\documentstyle[twocolumn,aps,prl,epsf,floats]{revtex}
\input epsf
\begin{document}
\draft
\date{18 June 97; printed \today}
\title{Are Steadily Moving Crystals Unstable?}
\author{Rangan Lahiri\cite{a} and Sriram Ramaswamy\cite{b}}
\address{Department of Physics, Indian Institute of Science,
Bangalore 560 012, India}
\maketitle
\begin{abstract}
We study the dynamics of small fluctuations about the uniform
state of a crystal moving through a dissipative
medium, e.g. a sedimenting colloidal crystal or a moving flux lattice,
using  a
set of continuum equations for the displacement fields, and
a one-dimensional driven lattice-gas model for the coupled
concentration and tilt fields.  
For the colloidal crystal we predict
a continuous nonequilibrium phase transition to a clumped state above
a critical P\'{e}clet number. 


\end{abstract}
What is the response of a crystalline lattice to a small, 
long-wavelength disturbance? Elastic theory \cite{ll} and its 
extension to time-dependent phenomena \cite{mpp} provide a complete 
answer to this question for a system {\em in thermal equilibrium}.  
In this Letter we ask and answer the same question 
for a lattice being {\em driven through a dissipative
medium} by a constant external force. This important  
nonequilibrium steady state arises, for example, in the steady sedimentation 
\cite{sed,rutgers} of a colloidal crystal \cite{aks} and in the 
motion of a depinned flux lattice in a current-carrying superconductor.   
Using symmetry arguments we construct continuum 
and lattice-gas models for the dynamics of small 
distortions about the uniform state of such a steadily moving lattice. 
The effects that we discuss arise from the dependence of the 
mobility of a given region on the local strain of the crystal. 
Our most striking result is that the dominant linear response at 
long wavelengths is proportional to the {\em driving speed} 
of the lattice, not to its elastic constants, and that this response 
can lead to a nonequilibrium phase transition.  

Before presenting our results in more detail let us recall 
an important  early study. Crowley \cite{crow} 
carried out experiments (on steel balls dropped gently into 
turpentine oil) and theory (calculating the hydrodynamic interactions 
between the spheres) to show that a regular horizontal array of 
sedimenting spheres was {\em linearly unstable} to clumping and buckling. 
Elastic forces, Brownian motion, and nonlinearities, all of which 
can resist this instability, were not considered in \cite{crow}. 
Since experiments on model colloidal systems are most conveniently 
performed by varying interaction strength rather than temperature 
\cite{aks}, the appropriate P\'{e}clet number for this problem is the 
ratio $Pe = \Delta \rho a g /G\/$ of gravitational to {\em elastic} rather  
than Brownian stresses. Here $\Delta \rho\/$ is the difference between 
the mass-densities of particles and solvent, $a\/$ the particle 
radius, $g\/$ the acceleration due to gravity, and $G\/$ a typical  
elastic modulus of the suspension. A sedimenting colloidal crystal, 
according to \cite{crow}, is therefore unstable  
in the $Pe = \infty\/$ limit. Hence the question posed in our title: 
does the instability persist at finite $Pe\/$?   

Our answer to this question is in two parts: (i) analytical results from a 
system of coupled nonlinear stochastic partial differential equations for 
displacement fields, and  
(ii) numerical studies of an equivalent driven lattice gas model \cite{ddlg} 
describing the coupled dynamics of concentration and tilt fields. 
Our simulations of the lattice model are done in the parameter 
range where there is a local tendency to undergo the 
Crowley \cite{crow} instability. We find that the system is nonetheless 
{\em stable} with respect to clumping upto a critical 
$Pe$ at which a {\em continuous nonequilibrium 
phase transition} to a clumped state takes place. For the 
corresponding parameter range the 
continuum model, neglecting nonlinearities and noise, is unstable 
without threshold, i.e. at $Pe = 0$. In view of the 
results of the lattice model, we expect that nonlinearities and fluctuations 
shift the critical $Pe$ for the onset of clumping in the 
continuum model to a nonzero value.  

We first derive the continuum model and 
perform a mean-field analysis, 
then show how the lattice-gas model follows naturally from it.  
Ignoring inertia altogether, which is justified for the experimental 
geometry we wish to consider \cite{fnlongrangehd}, the displacement field 
{\bf u}({\bf r}, t) of a $d$-dimensional lattice moving through a 
frictional medium 
with a mobility which depends on the local strain obeys  
\begin{equation}
\dot {\bf u} =   {\mbox{\boldmath $\mu$}}(\nabla {\bf u}) 
({\bf D}\nabla \nabla {\bf u} + {\bf F} + {\bf \zeta})
\label{eomgeneral}
\end{equation}
where 
the first term on the right represents, through the tensor ${\bf D}$,  
the elastic
restoring forces, the second is the driving force, and the third is 
a random force of thermal or possibly hydrodynamic origin \cite{brady}. 
{\boldmath $\mu$} is the mobility tensor which, 
in the single particle limit for colloids in a solvent with viscosity 
$\eta\/$, approaches the isotropic Stokes's Law value 
$\frac{1}{6 \pi \eta a}$.
We expand {\boldmath $\mu$} in powers of $\nabla {\bf u}$:
\begin{equation}
{\mbox {\boldmath $\mu$}} ({\bf \nabla u}) =
{\mbox {\boldmath $\mu_0$}} 
+ {\bf A } {\nabla \bf u} + 
{\cal O} (({\nabla {\bf u})}^2), 
\label{mobility}
\end{equation}
where {\boldmath $\mu_0$} is the 
mean macroscopic mobility of the undistorted crystal.  
For steady sedimentation along $z\/$, assuming isotropy in the $d-1$ 
transverse ($\bot$) dimensions but {\em not} 
under $z \rightarrow -z\/$, (\ref{eomgeneral}) and (\ref{mobility}) 
lead directly to 
\begin{mathletters}
\begin{eqnarray}
\dot {{\bf u}_{\bot}}& = &{\lambda}_1 {\partial}_z {\bf u}_{\bot} +
{\lambda}_2 {\nabla}_{\bot}  u_z
\nonumber
\\
&&+ {\cal O} ( \nabla \nabla u)
+ {\cal O} ( \nabla u  \nabla u) + {\bf f}_{\bot},
\label{eom1}
\\
\dot { u_{z}}& = &{\lambda}_3 {\nabla}_{\bot}.{\bf u}_{\bot} +
{\lambda}_4 {\partial}_{z}  u_z
\nonumber
\\
&&+ {\cal O} ( \nabla \nabla u)
+ {\cal O} ( \nabla u  \nabla u) +  f_z,
\label{eom2}
\end{eqnarray}
\label{eom}
\end{mathletters}
\noindent
where the constant drift along $z$ has been removed by comoving with
the crystal \cite{itp}.
Here ${\bf f}$ is a spatiotemporally white noise source \cite{mult}  
and ${\lambda}_i$ are phenomenological coefficients whose origin, in 
the case of a colloidal crystal, lies in the hydrodynamic interaction 
between the particles \cite{crow,fnlongrangehd,happel}. We explain the 
physical content of the terms in (\ref{eom}) below. First note that  
linearizing and Fourier transforming in space and time yields 
modes with frequencies of the form  
$\omega = \pm \sqrt{\lambda_2 \lambda_3} q - i D q^2$
for $q_z = 0, q_{\bot} \longrightarrow 0$.
If $\lambda_2 \lambda_3 > 0$, this leads to wavelike excitations  
at small wavenumber $q$, which are not the usual shear-waves 
of a crystal at equilibrium. The latter have been turned already into 
diffusive modes by the frictional dynamics adopted in (\ref{eom}). The speed 
of these waves is determined by the strain-dependence of the mobility, 
and the damping by the tensor $\bf{D}\/$ which is a ratio of elastic 
constants to friction coefficients. 

When $\lambda_2 \lambda_3 < 0$, 
the dispersion relation at small $q$ becomes $\omega \sim \pm i q$  
so that the model is linearly unstable with growth rate $ \propto q$. 
Thus there exist long-wavelength distortions of 
the perfect lattice which grow exponentially in time within the linear theory.
We do not know the sign of $\lambda_2 \lambda_3$ for moving
flux lattices, but for colloidal crystals it is negative, 
making them linearly unstable.
This is because hydrodynamic interactions cause denser regions
in the suspension to sink faster, and tilted regions to move laterally
in a way so as to cause an instability \cite{crow}.
Note that in (\ref{eom}) the {\em linear} elasticity of the crystal enters at 
second order in wavenumber and can thus not alter our conclusions 
about linear stability at long wavelengths.  
For $q > q_* \sim \sqrt{\lambda_2 \lambda_3} / D$, 
elastic forces suppress the linear instability.  
Small crystals are thus linearly stable.

To go beyond this linear analysis is daunting: even in $d = 2\/$, 
symmetry permits nine terms bilinear in $\nabla {\bf u}\/$ and 
six linear second derivative terms. Remarkably, all 
the essential physics is retained in a greatly simplified version 
in {\em one} space dimension. Consider a two-dimensional 
crystal described by a two-component displacement field (${u_x,u_z}$),
with the  sedimentation direction $z$ averaged out so that 
{\em only} $x\/$ {\em derivatives are included}. 
The equations of motion, retaining the lowest order nonlinearities 
and gradients, read  

\begin{mathletters}
\begin{eqnarray}
\dot{u_x}& = &{\lambda}_2 {{\partial}_x}u_z
+ {\gamma}_1 \partial_x u_x \partial_x u_z 
+ D_1 {{\partial}_x}^2 u_x +f_x
\label{eom1dconc}
\\
\dot{u_z} &=& {\lambda}_3 \partial_x u_x 
+ {\gamma}_2 (\partial_x u_x)^2 
\nonumber
\\
&&+ {\gamma}_3 (\partial_x u_z)^2 
+ D_2 {{\partial}_x}^2 u_z 
+f_z
\label{eom1dtilt}
\end{eqnarray}
\label{eom1d}
\end{mathletters}
The physics of each of the terms in (\ref{eom}) or (\ref{eom1d}) 
is reasonably clear. The first two terms on the right of 
(\ref{eom1dconc}) say that a tilt (i.e. $\partial_x u_z$)
produces a lateral drift. 
The first two terms on the right of (\ref{eom1dtilt}) contain the 
concentration dependence, and the third (a Burgers/KPZ-like term \cite{kpz}) 
the tilt-dependence, of the vertical sedimentation speed. 
Note that the $\gamma_i$ terms can be seen 
as arising from the dependence of the $\lambda_i$s on concentration and tilt .  
The second derivative terms in both equations arise simply from
elastic forces,  
and $f_x$ and $f_z$ are spatiotemporally white noises. 
The symmetry of (\ref{eom1d}) is under the {\em joint} inversion $x \rightarrow 
-x, \, u_x \rightarrow -u_x\/$. 

For $\frac{\gamma_1}{2 \gamma_2} = 
\frac{\lambda_2}{\lambda_3}\/$, 
$u_x \rightarrow u_x - \lambda_2 x / \gamma_1\/$ eliminates the $\lambda_2\/$ 
and $\lambda_3\/$ 
terms from (\ref{eom1d}), reducing it to the model of Erta\c{s} and 
Kardar (EK)   
\cite{ertas} in their limit $\lambda_{\perp}=\lambda_{\vert\vert}\/$, with 
its higher symmetry ($x \rightarrow -x\/$), 
albeit in a state of nonzero mean $\partial_x u_x\/$. The 
fluctuation-dissipation theorem, Galilean invariance and Cole-Hopf 
properties that arise in \cite{ertas} for special parameter values 
thus obtain here as well. 
If in addition ${\gamma}_1 = {2 \gamma}_3$
and $D_1 = D_2$,
the equations decouple in terms of transformed variables 
${\phi}_{\pm} = u_x \pm {\sqrt{\frac{\lambda_2}{\lambda_3}}} u_z$
into two equations: 
${\dot {\phi}_{\pm}} =
\pm {\sqrt {\lambda_2 \lambda_3}} \partial_x {{\phi}_{\pm}}
+D_1 {\partial_x}^2 {{\phi}_{\pm}}
\pm {\sqrt \frac{\lambda_2}{\lambda_3}} \gamma_2 {(\partial_x \phi_\pm) }^2
+f_{\pm}$,
a pair of independent KPZ \cite{kpz} equations 
with oppositely directed kinematic wave \cite{lighthill} terms.
and nonlinear couplings with opposite signs.
Clearly, for these parameter values, the stable driven  
crystal should exhibit KPZ exponents in its correlation functions.
The relevance of perturbations about the highly symmetric EK limit, 
as well as the statics and dynamics of the `stable' case 
$\lambda_2 \lambda_3 > 0\/$ will be studied in later work. 
In the present paper we focus on $\lambda_2 \lambda_3 < 0\/$.

Let us first look for steady-state 
solutions to (\ref{eom1d}) in the absence 
of noise, in terms of  $\rho = \partial_x u_x\/$ (the local 
concentration fluctuation) and $\theta = \partial_x u_z\/$ 
(the local up or down tilt). 
If we restrict ourselves for simplicity to spatially 
uniform states with left-right symmetry (so that $\theta = 0\/$, 
and the net currents of $\rho\/$ and $\theta\/$ are zero), we are 
left with only two possibilities: $\rho = \theta = 0\/$ or $\rho = -\lambda_3 / 
\gamma_2, \, \theta = 0\/$. In the vicinity of
$r \equiv \lambda_2 \lambda_3= 0\/$, the former
is stable for $r > 0\/$, the latter 
for  $r < 0\/$. This exchange of stabilities leads to a continuous onset 
of the $\rho \neq 0\/$ state, $\rho \sim |r|^{\beta}\/$ with $\beta = 1\/$. 
Similar analysis \cite{lahiri} gives a correlation length diverging 
as $|r|^{-\nu}\/$ with $\nu = 1/2\/$. 

Instead of attempting a perturbative treatment of the effect of 
nonlinearities and fluctuations on the above mean-field picture, 
we replace the continuous variable $x$ by a discrete index $i$, 
and $\partial u_x / \partial x$ by $\rho_i = u_x(i + 1) - u_x(i)$ 
[similarly $\theta_i = u_z(i+1) - u_z(i)$], with $\rho_i$ and $\theta_i$ 
restricted to $\pm 1$. Such an approach \cite{ddlg} 
has proved very successful for simulating the KPZ equation.  
The ``paramagnetic'' phase of these Ising variables
corresponds to the the undistorted crystal, and the  
``ferromagnetic'' phase represents
a macroscopically clumped and tilted state,
in terms of suitable order parameters which we define below. 
The best way to visualize the discrete model is to think of two sublattices:
a typical configuration can then be described by a sequence of spins 
$ {\rho}_1 {\theta}_1 {\rho}_2 {\theta}_2 {\rho}_3 {\theta}_3 \ldots$.
The dynamics of the spins is constructed by analogy with lattice 
models \cite{ddlg} for the KPZ equation,  retaining the essential 
features of (\ref{eom1d}), {\em viz.}, conservation of $\theta$ and 
$\rho$, stochasticity, lack of up-down symmetry, and the bias provided 
by each species on the motion of the other. The two approaches 
should yield identical long-distance properties.

Let us denote the states of $\rho_i$ by `$+$',`$-$'
and those of $\theta_i$ by `$/$' (uptilt) and `$\backslash$' (downtilt).
In the update rule corresponding to the linearly unstable case of 
(\ref{eom1d}), the rates for the following exchanges are enhanced relative to 
the corresponding reverse rates:
$+ \backslash - \rightarrow - \backslash +$, 
$-/+ \rightarrow +/-$, 
$/ + \backslash \rightarrow \backslash + /$ 
and 
$\backslash - / \rightarrow / - \backslash$ 
Since we are modelling charge-stabilised suspensions, it is 
useful to introduce a repulsion between regions of high density 
in the form of an enhanced probability for $++-$ or $-++$ to go to $+-+$.
Combining all of the above, we get the following exchange probabilities
for adjacent pairs of concentration and tilt:
\begin{mathletters} 
\begin{eqnarray} 
P_{{\rho}_i,{\rho}_{i+1}}& =& 
D_{\rho} - \epsilon_{\rho} {\theta}_i {\rho}_i 
+\alpha \{(1 + {\rho}_i)(1 + {\rho}_{i-1}) 
\nonumber \\
&&+ (1 + {\rho}_{i+1})(1 + {\rho}_{i+2}) \},
\label{latticemodel1}\\
P_{{\theta}_i,{\theta}_{i+1}} &=& D_{\theta} 
+ \epsilon_{\theta} {\rho}_{i+1} {\theta}_i
+ g_1 {\rho}_{i+1} 
+ g_2 {\theta}_i,
\label{latticemodel2}
\end{eqnarray}
\label{latticemodel}
\end{mathletters} 
\noindent
where $D_{\rho}$ and $D_{\theta}$ are related to the elastic constants,
${\epsilon}_{\rho}$, ${\epsilon}_{\theta}$, $g_1$ and $g_2$
to the $\gamma_i$ and $\lambda_i$ in (\ref{eom1d}), 
and $\alpha$ is the repulsion \cite{fnvalues}. 
Note that decreasing $\alpha$ reduces the stiffness of the system, 
thus increasing the effective P\'{e}clet number.  
The last two terms in (\ref{latticemodel2}) 
arise because of the lack of up-down symmetry.    
Our results in this paper are for $\epsilon_{\rho} \epsilon_{\theta} > 0\/$,  
corresponding to $\lambda_2 \lambda_3 < 0\/$ in (\ref{eom1d}). 

The mean value of both $\rho\/$ and $\theta\/$
are expected to be zero in the experimental system. 
We worked, therefore, at zero total `magnetization' for both  fields
and studied the model starting from random initial 
conditions, evolving it according to the above update rules for various 
system sizes $N\/$.
Periodic boundary conditions were used for all the runs.
For high values of repulsion
the spin configurations continued to be homogeneous under
time evolution.
When the repulsion was small or absent,
there was a phase separation 
into regions of high and low concentration and of up and down tilt,
separated by interfaces.
Thus the lattice seems to be stable for strong repulsion, 
but undergoes Crowley's clumping instability  \cite{crow} for weak repulsion.
The same behaviour, 
qualitatively, is observed as $D_{\rho}\/$ or $D_{\theta}\/$ 
are increased keeping other parameters fixed \cite{lahiri}.

\begin{figure}
\epsfxsize=8cm 
\epsfysize=6.5cm 
\centerline{\epsfbox{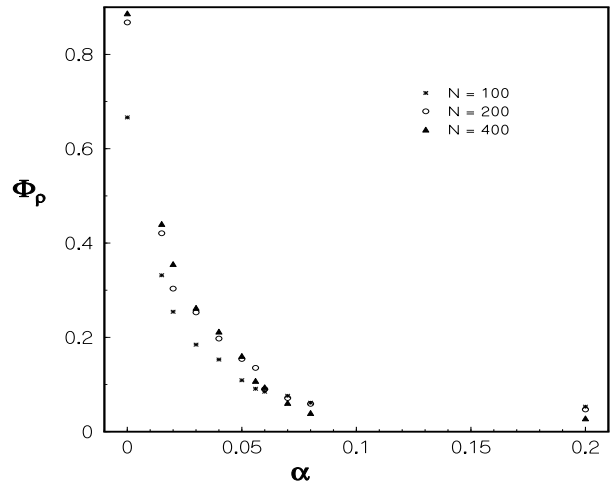}}
\caption[]{Order parameter (${\Phi}_{\rho}$) as a function 
of repulsion strength $\alpha$ for various system sizes N.
Increasing $\alpha$ is like decreasing P\'{e}clet number.}
\label{figorderparameter}
\end{figure}

To describe the ordered phase in this model with conserved dynamics,
we use two essentially equivalent order parameters:
${\Phi}_{\rho} = \sqrt{- \frac{1}{N} {\sum}_{i} {\rho}_i {\rho}_{i + N/2}}$ 
(similarly ${\Phi}_{\theta}$ for tilt)
which measure how anticorrelated the spins are across half the system size; 
and ($|\Psi_{\rho}|$, $|\Psi_{\theta}|$),
the moduli of the Fourier amplitudes of the spin fields
at the smallest nonzero wavevector $k_1 = \frac{2 \pi}{N}$
\cite{chandan}
(the amplitude at $k=0$ is zero).
Fig. \ref{figorderparameter} shows that the  
order parameter $\Phi_{\rho}\/$ is appreciable for small repulsion and 
decreases rapidly to a value consistent with zero for sufficiently large 
repulsion $\alpha$.  
Moreover, $\Phi_{\rho}\/$ {\em increases} with $N\/$ for $\alpha\/$ small 
and {\em decreases} \cite{lahiri} roughly as $1/\sqrt{N}\/$ for  $\alpha\/$ 
large.  
There must thus be a continuous nonequilibrium phase transition at 
$\alpha$ around .05, although to pin down the critical value 
of $\alpha\/$ would require careful finite-size scaling. 

We now present an independent check that the observed 
phase-separation is not merely the result of transients.
A  truly phase-separated state in a system of length $N$ should have
barriers to remixing which grow as $N^{\zeta}$ for some power $\zeta$.
The lifetime of such a state would then go as $\exp({N^\zeta})$.
To look for such barriers,
we define a lifetime $\tau (N)$ to be the mean time of first
passage of the order parameter $|\Psi|$ from a value 
$b_2 {|\Psi|}_{\mbox{\footnotesize max}}$
to $b_1 {|\Psi|}_{\mbox{\footnotesize max}}$, 
where  ${|\Psi|}_{\mbox{\footnotesize max}}$ is the maximum value of $|\Psi|$
over the runs.
$b_1$ and $b_2$ are numbers {\em independent} of N, chosen to get good 
statistics; we took $b_2 = 0.8, b_1 = 0.5$.

\begin{figure}
\epsfxsize=8cm 
\epsfysize=6cm 
\centerline{\epsfbox{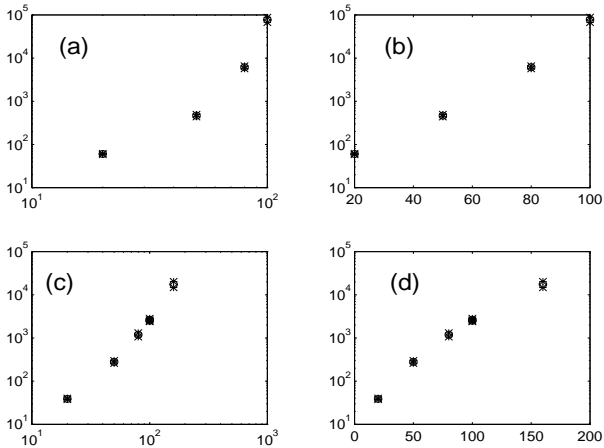}}
\caption[]{ Lifetime of the ordered phase $\tau (N)$, from the concentration
order parameter $|{\Psi}_{\rho}|$. 
$ \alpha = 0$: (a) log-log and (b) semi-log plot.
$ \alpha = 0.015$: (c) log-log and (d) semi-log plot.
Similar results hold for the tilt order parameter \cite{lahiri}.
Note the upward curvature of the log-log plots in either case, indicating
lifetimes growing faster than any power of $N$.
The downward curvature in (d) indicates that $\tau (N) $ is slower
than $e^N$ for $\alpha = .015$.
The error in $\tau$ is determined by allowing the system to run for several
lifetimes and statistically estimating the standard deviation.
}
\label{figexptimes}
\end{figure}

Fig. \ref{figexptimes} shows that $\tau (N)$
is roughly exponential in $N$ for $\alpha = 0$
and distinctly faster than a power law, presumably a stretched exponential,
for $\alpha = 0.015$.
This is strong evidence \cite{evans} 
for a true clumped phase at weak repulsion.
%Note that $\alpha = 0$ does not correspond
%to $Pe = \infty$, as fluctuations are present through
%$D_{\rho}$ and $D_{\theta}$.

To see why phase separation can occur in this one-dimensional model 
one has to look at the positions of the concentration and tilt domains.
We find in our simulations that the system goes into a steady state
in which the domains are
staggered with respect to each other by an approximate distance $N/4$.
This happens in such a way that a concentration interface 
$+++---$lives in a region crowded with uptilts $/$ which inhibit the 
exchange of a pair $+-$.
The dissolution of the interface by interdiffusion of $+$ and $-$
thus requires uphill motion over a nonzero fraction of $N$.

Since our simple one-dimensional model undergoes a clumping transition,
it is reasonable to expect that a real charge-stabilized colloidal
crystal in a fluidized bed \cite{rutgers,fnlongrangehd} will do so as well.
The repulsion between polyballs may be decreased by adding salt to
the fluid, which should lead to an observable clumping transition at 
ionic strengths much lower than those required to produce
melting or aggregation at equilibrium.
The clumping will manifest itself
as a breakup of the crystal into smaller crystallites
(since the crystal is stable at small enough system size),
separated by regions of strong upward fluid flow  \cite{poon}.
A detailed analysis of this dynamics would require the inclusion
of the hydrodynamic flow.

In summary, we have demonstrated that the long wavelength 
dynamics of a crystal moving steadily through a dissipative 
medium is qualitatively different from its equilibrium 
counterpart. In particular, we have shown that a natural 
driven lattice gas model for this system shows a dramatic 
nonequilibrium phase transition to a clumped state, and 
we urge experimenters to test our predictions. 

We thank 
M. Barma, P. M. Chaikin, C. Dasgupta, D. Dhar, D. Erta\c{s}, 
G. Fredrickson, M. Kardar, D. Mukamel, A. Pande, R. Pandit, F. Pincus, 
W.C.-K. Poon, A. Sain, and B. S. Shastry for suggestions,
M. A. Rutgers for his thesis,  
and C. Das, R. Govindarajan and P. Pradhan for help 
with computing.  
This work began in discussions by SR with   
L. Balents, M.P.A. Fisher, and M.C. Marchetti at the ITP 
Biomembranes Workshop (NSF grant PHY94-07194), and was done in part at 
the Isaac Newton Institute (NATO ASI and Workshop
on the Dynamics  of Complex Fluids).  
We thank SERC, IISc Bangalore for computational resources,  
and CSIR and BRNS for financial support.


\begin{references}
\bibitem{mult} 
For simplicity we have ignored multiplicative noise terms.
\bibitem{fnlongrangehd}
Our description, based on assumptions of locality and gradient expansion,
neglects the long-range nature of hydrodynamic interactions.
The experimental situation that corresponds to this is a 
two-dimensional colloidal crystal confined between parallel walls
(aligned with the $xz$ plane),  
which cut off the hydrodynamic forces, making
them effectively short-ranged.
\end{references}
\end{document}